\documentclass{elsart}
\usepackage{natbib}
\usepackage{epsfig}
\begin{document}
\begin{frontmatter}
\title{Physics Case and Challenges for\\ the Vertex Tracker at\\ 
Future High Energy $e^+e^-$ Linear Colliders}
\vspace*{-0.30cm}
\author{M.~Battaglia}
\address{CERN, CH-1211 Geneva 23, Switzerland}
%

\begin{abstract}
The physics programme of high energy $e^+ e^-$ linear colliders relies on 
the accurate identification of fermions in order to study in details the 
profile of the Higgs boson, search for new particles and probe the multi-TeV
mass region by means of precise electro-weak measurements and direct searches.
\end{abstract}
\end{frontmatter}
\vspace*{-1.0cm}
\section{Introduction}

\vspace{-0.5cm}

With the end of the LEP program and the start of operation of the 
$B$-factories, the next step in experimentation at lepton colliders will be
at linear accelerators (LC) able to deliver $e^+e^-$ collisions at 
centre-of-mass energies $\sqrt{s}$ in the range 0.3-1.0~TeV with a 
luminosity of the order of $10^{34}$cm$^{-2}$s$^{-1}$ and eventually to 
achieve $\sqrt{s}$ = 3-5~TeV with luminosity of 
$10^{35}$cm$^{-2}$s$^{-1}$ or higher.
The lower end of this energy range will be devoted to the accurate study of 
the Higgs profile, if a relatively light Higgs boson exists as suggested by 
the LEP and SLC data, to the precise determination of the top mass and to 
complement the LHC program in searching for signals of new physics beyond the 
Standard Model. A multi-TeV linear collider is expected to break new grounds, 
by exploring the mass scale beyond 10~TeV for new phenomena and to study in 
detail the properties of the new physics, established by the LHC or the 
first LC phase.

Present projects focus on three different beam acceleration techniques:
warm RF (mostly X-band as for the NLC~\cite{nlc} and JLC~\cite{jlc} projects)
or super-conducting cavities (such as TESLA~\cite{tesla}) and two-beam 
acceleration at high frequency (such as CLIC~\cite{clic}). 
The proposed technologies
being optimised for different $\sqrt{s}$ energies, the full energy range can be
covered by subsequent upgrades using the same infrastructures or by planning 
two distinct phases corresponding to different accelerators.

The Vertex Tracker of the LC detector is expected to provide the LC data with 
the fermion flavour tagging 
capabilities that are instrumental to the understanding of the electro-weak 
symmetry breaking mechanism and in the search for new particles. 
Several Vertex Tracker designs have been proposed, relying on different sensor
technologies. A common denominator of these proposals is the use of pixel 
devices due to the high particle density and the emphasis on the material 
budget minimisation to improve the track extrapolation performances at small 
momenta in multi-jet final states. 

\vspace{-0.75cm}

\section{Experimental Conditions for the Vertex Tracker}

\vspace*{-0.5cm}
The Vertex Tracker at the linear collider will be exposed to background 
and radiation levels and to track densities unprecedented for $e^+e^-$ 
collider machines. The main source of background in the interaction region 
is due to $e^+e^-$ pairs produced and bent in the intense electro-magnetic 
interaction of the colliding beams. The hit densities have been estimated 
for each collider design using dedicated beam simulation programs and tracking
of the produced particles in the detector solenoidal field. 
The radius and maximum length of the innermost sensitive layer are defined by 
the envelope of the deflected particles from pairs. Results are summarised
in Table~1.

\begin{table}[h!]
\begin{center}
\caption{Pair hit density on innermost Vertex Tracker layer
for different LC designs and $\sqrt{s}$.}
\begin{tabular}{l c c r r r}
\hline
LC & $\sqrt{s}$ (TeV) & R (cm) & Hits mm$^{-2}$ BX$^{-1}$ & 
25 ns$^{-1}$ & train$^{-1}$\\
\hline \hline
NLC   & ~0.5 & 1.2 & 0.100  & 1.80 & 9.5 \\
TESLA & ~0.8 & 1.5 & 0.050  & 0.05 & 225.0 \\
CLIC  & ~3.0 & 3.0 & 0.005  & 0.18 & 0.8 \\ \hline
\end{tabular}
\end{center}
\label{tab:pairs}
\end{table}
While the pair background sets the most stringent constraint on the Vertex 
Tracker geometry, defining an inward bound on the detector radial position,
track density in highly collimated hadronic jets contributes significantly to 
the local detector occupancy. 
Typical values for the local track densities are expected to be of the 
order of 0.2-1.0 hit~mm$^{-2}$ at $\sqrt{s}$ = 0.5~TeV, depending on the 
physics process, and about a factor three larger at $\sqrt{s}$ = 3~TeV for 
solenoidal fields of 4~T and 6~T respectively. Values for the anticipated 
total hit density are comparable to, or larger than, those foreseen on the 
innermost vertex detector layers at the LHC, ranging between 
$\simeq$~0.03~hits/25~ns for ATLAS and 0.9~hits/25~ns in ALICE.

Other background sources may also be critical for the sensor design.  
Neutrons are copiously produced by the dump of pairs, beamstrahlung photons and
spent beams. Their flux into the sensitive detector volume must be reduced
by a proper choice of surface coatings and by isolating the detector from
the accelerator tunnel. Detailed simulation of neutron production and 
transport show that the expected neutron fluxes are in the range of few times
10$^9$ 1~MeV-equivalent neutrons cm$^{-2}$ year$^{-1}$ for the NLC and TESLA
and about an order of magnitude larger at a multi-TeV collider. While these 
fluxes are several orders of magnitude smaller compared to those expected at 
the LHC, they still impose special care for some of the proposed sensor 
technologies, such as Charged Coupled Devices (CCD), to minimise the 
inefficiencies arising from the creation of charge trapping sites due to bulk 
radiation damage.
Finally $\gamma \gamma \rightarrow {\mathrm{hadrons}}$ events overlayed
on a $e^+e^-$ interactions need to be minimised. In fact, the additional 
visible energy and the displaced particle vertex from $\gamma \gamma 
\rightarrow {\mathrm{hadrons}}$ may distort the event reconstruction in 
channels with missing energy signatures and affect the jet flavour tagging 
performances~\cite{satoru}. 
LC designs with large luminosity per bunch crossing or high 
$\sqrt{s}$ energy are more affected by the higher probability of a 
$\gamma \gamma$ event overlap. At $\sqrt{s}$ = 350~GeV, this probability is 
$\simeq 0.02$ BX$^{-1}$ and $\simeq 0.12$ BX$^{-1}$ in the case of NLC and 
TESLA respectively and becomes $\simeq 4.0$ BX$^{-1}$ at $\sqrt{s}$ = 3~TeV
for the CLIC parameters. Fast time stamping, providing bunch identification
information, will allow to reduce the total $\gamma \gamma$ overlap 
probability.

These issues emphasise the case for sensors with small pixel cells
and fast detector read-out or fast time stamping capabilities. 
Setting an acceptable rate 
of background as 5 hits mm$^{-2}$ and $< 0.5$ $\gamma \gamma$ event for 
precision physics at $\sqrt{s}$ = 350-500~GeV, results in the requirement to
identify hits from events within 70~ns and 1~$\mu$s for the NLC and TESLA 
projects. This requirement is well within the planned timing performances of 
the LHC pixel detector electronics and is being address by recent developments
in CCD read-out~\cite{ccd}.

\vspace{-0.50cm}

\section{Physics Objectives for the Vertex Tracker}

\vspace{-0.5cm}

The linear collider has the potential to cover the widest energy range for any
particle accelerator operated so far, ranging from the $Z^0$ peak up to the
multi-TeV region. Within this range, its physics programme covers 
precision electro-weak tests, a detailed study of the electro-weak symmetry
breaking mechanism and the search for new physics at, and beyond, the energy
scale probed by the LHC. This programme relies on efficient flavour 
identification to perform detailed studies of particle couplings to up- and
down-type quarks and to different fermion generations, aiming to probe the 
Higgs mechanism of mass generation and new phenomena such as Supersymmetry and
extra-dimensions, and to identify fermions of the heaviest generation 
($\tau$, $b$ and $t$) for searching for new particles, such as the heavy part 
of the Higgs sector in extensions of the Standard Model ($H^{\pm}$, $H^0$, 
$A^0$). Jet flavour tagging is based on the accurately measured track 
impact parameters, complemented by the topology of detected secondary and 
tertiary vertices and kinematical variables sensitive to the secondary track 
multiplicity and invariant mass of tau leptons and charm or beauty hadrons.

In addressing this variety of processes, the anticipated S/B ratios and 
kinematical conditions vary quite considerably. At energies around the $Z^0$ 
peak the typical decay length in space of a $B$ hadron is 
0.3~cm and about one in five $Z^0$ decays consists of a $b \bar b$ pair. The 
availability of data samples of the order of 10$^9$-10$^{10}$ $Z^0$ can 
improve the accuracy on $\Gamma_{b \bar b}$ and $A_b$ by factors of five and 
ten respectively, provided the $b$ tagging purity can be improved from 98\%, 
obtained by DELPHI, to 99.3\% for a constant efficiency of 30\%~\cite{moenig}. 
The investigation of CP violation in the 
$B$~sector and of rare decays at the $B$-factory and $pp$ collider 
experiments, may be valuably complemented by studies profiting of the unique 
kinematical properties of $Z^0 \rightarrow b \bar b$ decays.

At $\sqrt{s}$ = 0.3-0.8~TeV, $b$-tagging is instrumental in isolating a 
clean Higgs signal, for Higgs masses up to about 150~GeV/$c^2$. The study of 
the Higgs boson (mass, width, quantum number and couplings) in the
$e^+e^- \rightarrow HZ$ and $H \nu \bar \nu$ production processes needs a 
good rejection of the dominant $WW$ and $ZZ$ backgrounds, that can be achieved
by selecting $H \rightarrow b \bar b$ decays with high efficiency.
In the challenging Higgs-strahlung off top quark $t \bar t H$ and double Higgs
production, $HHZ$ and $HH \nu \bar \nu$, reactions, backgrounds need to be
rejected by almost five orders of magnitude in order to measure the top Higgs
Yukawa coupling and the triple Higgs coupling, necessary to complete the 
investigation of the Higgs coupling properties.
Four $b$-jets, in multi-jet final states from $t \bar t H \rightarrow 
b W^+ \bar b W^- b \bar b$, $HHZ \rightarrow b \bar b b \bar b q \bar q$ and 
also Supersymmetric charged Higgs decays $H^+H^- \rightarrow t \bar b \bar t b
\rightarrow b W^+ \bar b \bar b W^- b$, offer a spectacular signature, while 
requiring high $b$-tagging efficiency due to the small production 
cross-section, corresponding to as few as 15 events per 100~fb$^{-1}$.
At the same time, the large $H \rightarrow b \bar b$
yield poses a major challenge to the extraction of the tiny branching 
fractions for $H \rightarrow c \bar c$ and $H \rightarrow g g$ decays that are
important as proof that the Higgs mechanism is responsible for the fermion mass
generation and to identify the Standard Model or Supersymmetric nature of the 
neutral Higgs boson~\cite{mb}. This requires charm tagging performing 
efficiently in presence of both beauty and light quark backgrounds. This
indicates the need of preserving an accurate track extrapolation resolution 
down to low momenta to tell the charm on the basis of the reconstructed mass 
and decay multiplicity of the detected short-lived hadron. Charm
tagging is also instrumental in the study of triple gauge boson couplings that
can be performed by analysing the reaction $e^+e^- \rightarrow W^+W^- 
\rightarrow \ell^+ \nu q \bar q^{'}$ at high energy. The capability of 
tagging charm jets with 85\% efficiency corresponds to a 50\% equivalent 
luminosity gain compared to flavour blind detector results, where information
is lost in the averaging of the $q \bar q^{'}$ states~\cite{walkowiak}.
$\tau$ leptons are expected to be a main signature for decay of Supersymmetric
particles in models with large values of the $\tan \beta$ parameter. Their 
detection in jets requires to match the impact parameter tag with the 
calorimetric response.

While the developments of Vertex Trackers from the LEP and SLD experiments
to those at B-factories, the Tevatron and the LHC have addressed 
many crucial aspects, such as radiation hardness, read-out speed and 
the considerable size increase of the trackers, the track extrapolation 
resolution, $\sigma_{ip}$, has not benefited of significant improvements. 
Limited in terms of multiple scattering the $B$ factories and by radiation 
damage in their closest approach to the beam the hadron colliders, their 
typical performances are comparable to those achieved at 
LEP, $\sigma_{ip}$ = 25~$\mu$m$\oplus$70~$\mu$m/$p_t$~GeV/$c$. 
In order to fulfill its challenging requirements, the LC Vertex Tracker aims 
at significantly improving on the already outstanding SLD VXD3 performances of 
8~$\mu$m$\oplus$33~$\mu$m/$p_t$ GeV/$c$~\cite{vxd3}. 
An improvement to $\sigma_{ip}$ to  5~$\mu$m$\oplus$10~$\mu$m/$p_t$ GeV/$c$ 
corresponds to a factor $\simeq~2$ increase in charm jet tagging efficiency 
and a factor $\simeq~1.25$ increase in beauty jet tagging efficiency at 
constant mis-identification probability.   
Since for a given detector geometry, $\sigma_{ip}$ depends on the sensor
thickness and single point resolution, these requirements have motivated a
dedicated R\&D activity to improve the pixel detector performances beyond
what already achieved for the LHC experiments~\cite{pixel} and a study of a 
new concept of unsupported CCD sensors to minimise the material 
budget~\cite{ccd}.

At multi-TeV energies, flavour dependent electro-weak observables, such as
$\Gamma_{c \bar c}$, $\Gamma_{b \bar b}$ and $A_{fb}^{b \bar b}$, offer a 
window on new phenomena, such as new gauge bosons, compactified 
extra-dimensions and contact interactions,  well beyond 10~TeV. The large 
$B$ hadron boost causes decay vertices to be located well beyond the first 
layers of the tracker and the decay products to be highly collimated. 
In $e^+e^- \rightarrow b \bar b$, the average $B$ decay length 
increases to $\simeq$10~cm with a local track density $>1$~mm$^{-2}$, thus 
requiring a large Vertex Tracker volume and, possibly, new jet flavour tagging
methods extending the perspectives for developments of new generations of 
Vertex Trackers at $e^+e^-$ colliders.

\end{document}